\documentclass[namedreferences]{solarphysics}
\usepackage[optionalrh,solaromanenum]{spr-sola-addons} 
\usepackage{graphicx}                    
\usepackage{color}                       

\begin{document}
\begin{article}
\begin{opening}

\title{Dependence of the Sunspot-group Size on the Level of Solar Activity and its Influence on the Calibration of Solar Observers}

%
\author{I.G.~\surname{Usoskin}$^{1,2}$ \sep
    G.A.~\surname{Kovaltsov} $^{3,1^\star}$\sep
    T.~\surname{Chatzistergos}$^{4}$\sep
    }

%
\runningauthor{I.G. Usoskin \textit{et al.}}
\runningtitle{Sunspot-group Size vs. Solar Activity}

%
\institute{ $^1${ReSoLVE, Space Climate Group, University of Oulu, Finland.}\\
 $^2${Sodankyl\"a Geophysical Observatory, University of Oulu, Finland.}\\
 $^3${Ioffe Physical-Technical Institute, St. Petersburg, Russia.}\\
 $^4${Max-Planck-Institut f\"ur Sonnensystemforschung, G\"ottingen, Germany}\\
 $^\star${(visiting scientist)}\\
                     email: {Ilya.Usoskin@oulu.fi}
             }

\begin{abstract}
The distribution of the sunspot group size (area) and its dependence on the level of solar activity is studied.
It is shown that the fraction of small groups is not constant but decreases with the level of solar activity
 so that high solar activity is largely defined by big groups.
We study the possible influence of solar activity on the ability of a realistic observer to see and
 report the daily number of sunspot groups.
It is shown that the relation between the number of sunspot groups as seen by different
 observers with different observational acuity thresholds is strongly non-linear and cannot be approximated by the
 traditionally used linear scaling ($k-$factors).
The observational acuity threshold [$A_{\rm th}$] is considered to quantify the quality of each observer, instead of
 the traditional relative $k-$factor.
A nonlinear $c-$factor based on $A_{\rm th}$ is proposed, which can be used to correct each observer to the reference conditions.
The method is tested on a pair of principal solar observers, Wolf and Wolfer, and it is shown that the traditional
 linear correction, with the constant $k-$factor of 1.66 to scale Wolf to Wolfer, leads to an overestimate
 of solar activity around solar maxima.
\end{abstract}


\end{opening}

\section{Introduction}
\label{Sec:Intro}
The sunspot number series was introduced in the 1860s by Rudolf Wolf of Z\"urich and became the most commonly used index of
 long-term solar variability ever since.
The sunspot number series is longer than 400 years, including the Maunder minimum (\opencite{eddy76,sokoloff04,usoskin_MM_15}),
 and is composed of {observations} from a large number of different observers.
Since they used different instruments and different techniques for observing and recording sunspots, it is
 unavoidable that data from different observers need to be calibrated to each other to produce a homogeneous dataset.
The first inter-calibration of data from different observers was performed by Rudolf Wolf in the mid-19th century.
He proposed a simple linear scaling between the different observers (the so-called $k-$factors) so that
 the data (count of groups and sunspots) from one observer should be multiplied by a $k-$factor to rescale it
 to another reference observer.
The value of the correction $k-$factor is assumed to be rigidly fixed, as found by a linear regression, for each observer,
 and it characterizes the observer's quality in a relative way with respect to the reference observer.
Since then, this method has always been used until very recently (\opencite{clette14,svalgaard16}).

The $k-$factor approach utilizes the {method of ordinary linear least square regression forced through the origin}.
This method is based on several formal assumptions which are usually not discussed, but their violation may
 lead to incorrect results:
\begin{enumerate}
\item
{\it Linearity}, \textit{i.e.} the relation between two variables $X$ and $Y$ can be described
 as linear in the entire range of the $X$-values.
 This assumption is invalid for the sunspot (group) numbers, as shown by \inlinecite{lockwood_SP3_2016} or \inlinecite{usoskin_ADF_16}
 and discussed here, because of the essential nonlinearity.
\item
{\it Random sample}, \textit{i.e.} the pairs of $X$- and $Y$-values are taken randomly from the same population and
 have sufficient lengths.
 This assumption is valid in this case.
\item
{\it Zero conditional mean}, \textit{i.e.} normality  and independence of errors, implying that all errors are normally
 distributed around the true values.
 This {assumption} is also invalid since the errors are asymmetric and not normal
 (\opencite{usoskin_ADF_16}).
\item
{\it Constant variance (homoscedasticity)}.
 This assumption is violated since the variance of the data is not constant but depends on the
 level of solar activity so that the variance of the data points is much larger for periods of high activity
 than around solar minima.
\item
X-values are supposed to be known exactly without errors.
This assumption is invalid since data from the calibrated observer
 (X-axis) can be even more uncertain than those by the reference observer (Y-axis).
\item
Additionally, forcing through the origin is assumed for the $k-$factors.
This assumption is also invalid as shown by
 \inlinecite{lockwood_SP3_2016}, since no spot reported by an observer does not necessarily mean that an observer
 with a better instrument would not see some small spots.
\end{enumerate}
We do not discuss here the issue of collinearity, since this assumption is not directly applied to the regression problem considered here.
Accordingly, five out of six assumptions {listed} above are invalid in the case of sunspot numbers making the linear scaling calibration
 by $k-$factor formally invalid.
This method was reasonable in the mid-19th century {for interpolations to fill short gaps in observations},
 but now we aim to develop a more appropriate method for a direct
 calibration of different observers to each other.
Several indirect methods of solar-observer calibration have been introduced recently (\opencite{friedli16,usoskin_ADF_16})
 but here we focus on a direct inter-calibration {based on} modern statistical methods.
\begin{figure}
\centering \resizebox{10cm}{!}{\includegraphics[bb = 28 177 486 805]{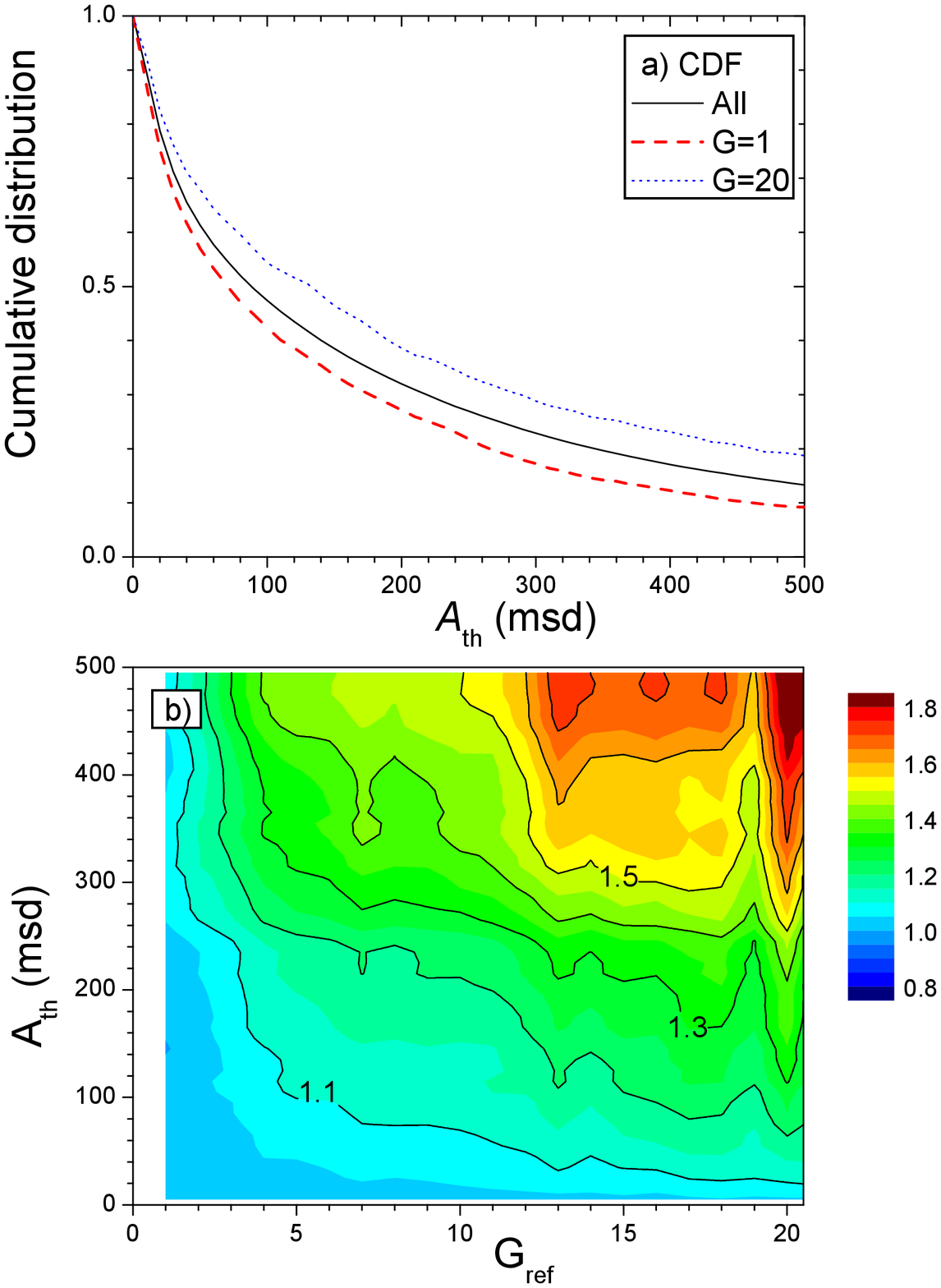}}
\caption{
Size distribution of sunspot groups from the reference database.
Panel a: Cumulative distribution function (CDF) of sizes of sunspot groups above the given threshold $A_{\rm th}$.
The solid line depicts the entire population of sunspot groups, the red dashed and blue dotted lines show the
 CDF for low-activity ($G=1$) and high-activity days ($G=20$), respectively.
Panel b: The 2D map of the CDF (color code is shown on the right) as a function of the activity level (the daily
 $G_{\rm ref}$ -- X-axis) and the group size (Y-axis), normalized to the CDF at $G=1$ (the red dashed
 curve in panel a).
}
\label{Fig:map}
\end{figure}

It was proposed recently (\opencite{lockwood_SP3_2016,usoskin_ADF_16}) that the ``quality'' of a solar observer
 can be quantified not by a relative $k-$factor but by the observational acuity threshold, \textit{i.e.} the
 minimum size of a sunspot group the observer can see considering the used instrumentation, technique and eyesight.
This quantity [$A_{\rm th}$] (in millions of the solar disc, msd) has a clear meaning -- all sunspot groups
 bigger than $A_{\rm th}$ are reported
 while all the groups smaller than $A_{\rm th}$ are missed by the observer.
We note that weather conditions and age or experience may lead to variations of the actual
 threshold for a given observer in time, but here we consider that the threshold is constant in time.
This is also assumed in the $k-$factor methodology.
The threshold would be consistent with the $k-$factor if the fraction of small ($<A_{\rm th}$) groups
 on the solar disc
 was roughly constant and independent on the level of solar activity.
However, as many studies imply (\opencite{kilcik11,jiang11_2,nagovitsyn12,obridko14}),
 the fraction of small groups varies with solar
 activity: it is large around solar minima and decreases with the level of solar activity.
Accordingly, the use of the linear $k-$factor method may lead to a distortion of the calibrated sunspot numbers
 (\opencite{lockwood_ApJ_16}).

In this article we study the relation between sunspot group counts by a ``poor'' observer and those by the
 reference ``perfect'' observer,
 using the reference dataset described in Section~\ref{Sec:data}.
In Section~\ref{Sec:dist} we study the distribution of sunspot group sizes and its dependence on the level of solar activity.
Its effect on observations by solar observers of different quality and their inter-calibrations are
 discussed in Section~\ref{Sec:obs}.
We propose the use of a new nonlinear $c-$factor to calibrate data from a ``poor'' observer to the
 reference conditions depending on the level of solar activity, in a more realistic manner than
 that offered by the traditionally used linear $k-$factor.

\section{Data}
\label{Sec:data}

We base our analysis on the Royal Greenwich Observatory (RGO)\footnote{http://solar\-science.msfc.nasa.gov/greenwch.shtml}
 data series of sunspot groups with their areas.
This series is referred to as the reference dataset throughout this article.
Although the RGO data series starts in 1874, there are indications that its quality might be variable before 1900
 (\opencite{clette14})
 {or even before 1915 (\opencite{cliver16})}
 due to the ``learning'' curve, although other studies did not find this effect or
 attributed it only to the very early part of the RGO record before 1880 (\opencite{sarychev09,carrasco13,aparicio14}).
To stay on the conservative safe-side, we consider here RGO data only for the period {1916--1976}
 when the data series is homogeneous in quality.
We have checked that the result remains qualitatively the same if the
 period of {1874--1915} is included into the analysis.
Since the RGO series was terminated in 1976, we also stop our reference dataset at that time,
 not extending it with data from the \textit{Solar Observing Optical Network} (SOON)
 because of a possible transition inhomogeneity (\opencite{lockwood_1_14,hathawayLR}).

To make the results compatible with the direct observations, we used in our analysis the uncorrected
 (for foreshortening) whole area of sunspot groups, \textit{i.e.} as it is seen from Earth.
We consider sunspot groups, not individual spots, since several closely located spots indistinguishable by the
 observer can be seen as one blurred spot
 even with a poor telescope and thus are more representative for the actual data.

From the reference RGO data series we compiled files of daily numbers of sunspot groups with different
 observational acuity thresholds [$A_{\rm th}$] quantified as the group areas in msd,
 so that a group is counted (observed) if its uncorrected total area is not smaller that the threshold
 value in msd.
The corresponding daily number of groups is denoted as $G_{\rm A}(t)$ where the subscript $A$ denotes
 the value of the threshold in msd.
For example $G_{100}$ denotes the number of groups with area $\geq 100$ msd for each day.
The number of groups without applying any threshold ($A_{\rm th}=0$, \textit{i.e.} the total number
 of groups in the reference
 dataset irrespectively of their size) is called the reference series $G_{\rm ref}$.

\begin{figure}
\centering \resizebox{12cm}{!}{\includegraphics[bb= 70 500 409 737]{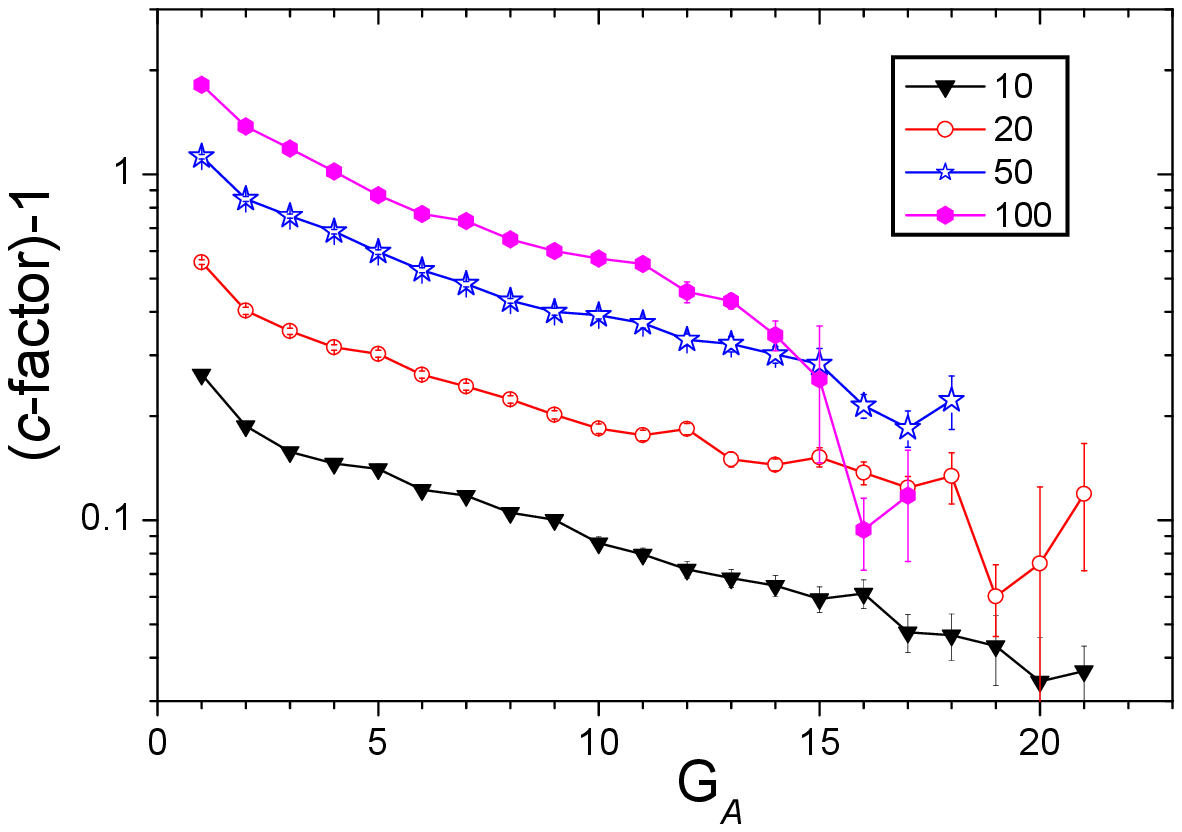}}
\caption{
Ratio of the ``small'' (\textit{viz.} smaller than the given value of the threshold [$A_{\rm th}$] in units of
 msd as denoted in the legend) to
 ``large'' sunspot groups as a function of the number of `large' groups, in the total daily number of sunspot groups for the RGO
 dataset for the period {1916--1976}.
 This corresponds to the correction $c-$factor (see text) minus 1.
 The error bars represent the standard statistical errors of the ratio.
 }
\label{Fig:CF}
\end{figure}

For the analysis of data by Rudolf Wolf and Alfred Wolfer (Section~\ref{Sec:WW}) we used the daily number of
 sunspot groups as presented in the
 database\footnote{http://www.ngdc.noaa.gov/stp/space-weather/solar-data/solar-indices/sunspot-numbers/group/}
 of \inlinecite{hoyt98} {and the new revised collection of
 sunspot group numbers\footnote{http://haso.unex.es/?q=content/data} by \inlinecite{vaquero16}, for
  the period of their overlap 1876-1893.
However, the latest (not available for the time of making the present analysis) revision of the Wolf's records
 (\opencite{friedli16_2}) is not included in these two databases.
On the other hand, the analysis of Wolf and Wolfer data is shown for illustration and would not be altered
 with the slightly revised dataset.}

\section{Distribution of Sunspot-group Sizes}
\label{Sec:dist}

Here we investigated how size of sunspot groups changes with solar activity.
This is usually studied using the mean size of sunspot groups (\opencite{jiang11_2}) but this may be confusing
 because the size distribution of spots is highly asymmetric and the {mean} value is not a robust feature.

Figure~\ref{Fig:map}a depicts the cumulative distribution function (CDF) of sunspot-group sizes (uncorrected
 total area in msd) for the reference dataset (Section~\ref{Sec:data}).
CDF($A_{\rm th}$) is defined as the fraction of the sunspot groups with area not smaller than the given value of the
 area threshold [$A_{\rm th}$].
By definition CDF(0) is equal to unity (each group has a non-zero size).
The black curve shows the global CDF for all the sunspot groups (about {119300} groups) in the reference dataset.
{The blue dotted line represents the CDF for 920 groups (46 days) for high activity days with 20 groups ($G=20$)
 reported.}
The red dashed curve depicts the CDF for {2781} days with low activity (only one group reported, $G=1$).
One can see that there is a significant fraction of large groups even for low-activity days:
 $\approx$10\% of groups have area greater than 500 msd.
The group-size distribution changes with the level of solar activity: while the CDF for low-activity days
 is lower than the global CDF, the distribution for high-activity days is significantly higher.
For example, as one can see from Figure~\ref{Fig:map}a, the relative contribution of large sunspot groups
 ($A\geq 500$ msd) doubles for high-activity days ($G=20$) with respect to low-activity days ($G=1$).
This implies that the rise of activity is mostly due to the emergence of large sunspot groups,
 indicating {that the sunspot-group size distribution changes with the level of solar activity}.

To generalize the study of the CDF dependence on the level of solar activity,
 we plotted in Figure~\ref{Fig:map}b a contour plot of the CDF as a function of the activity level (quantified in $G$)
 and the group size (quantified in msd).
All the CDF were normalized to that at the low-activity level ($G=1$, see red dashed line in Figure~\ref{Fig:map}a)
 so that CDF($G=1$) is unity for all $A_{\rm th}$.
One can see that the shape of CDF changes with the level of solar activity so that the fraction of large spots
 is growing, while the fraction of small spot decreases with activity.
For example, the brown spot in the top-right corner implies that the relative fraction of groups with area $>500$ msd
 is nearly doubled for high-activity days ($G=20$) compared to low-activity days ($G=1$), as discussed above
 for Figure~\ref{Fig:map}a.

\begin{figure}
\centering \resizebox{8cm}{!}{\includegraphics[bb = 62 205 330 739]{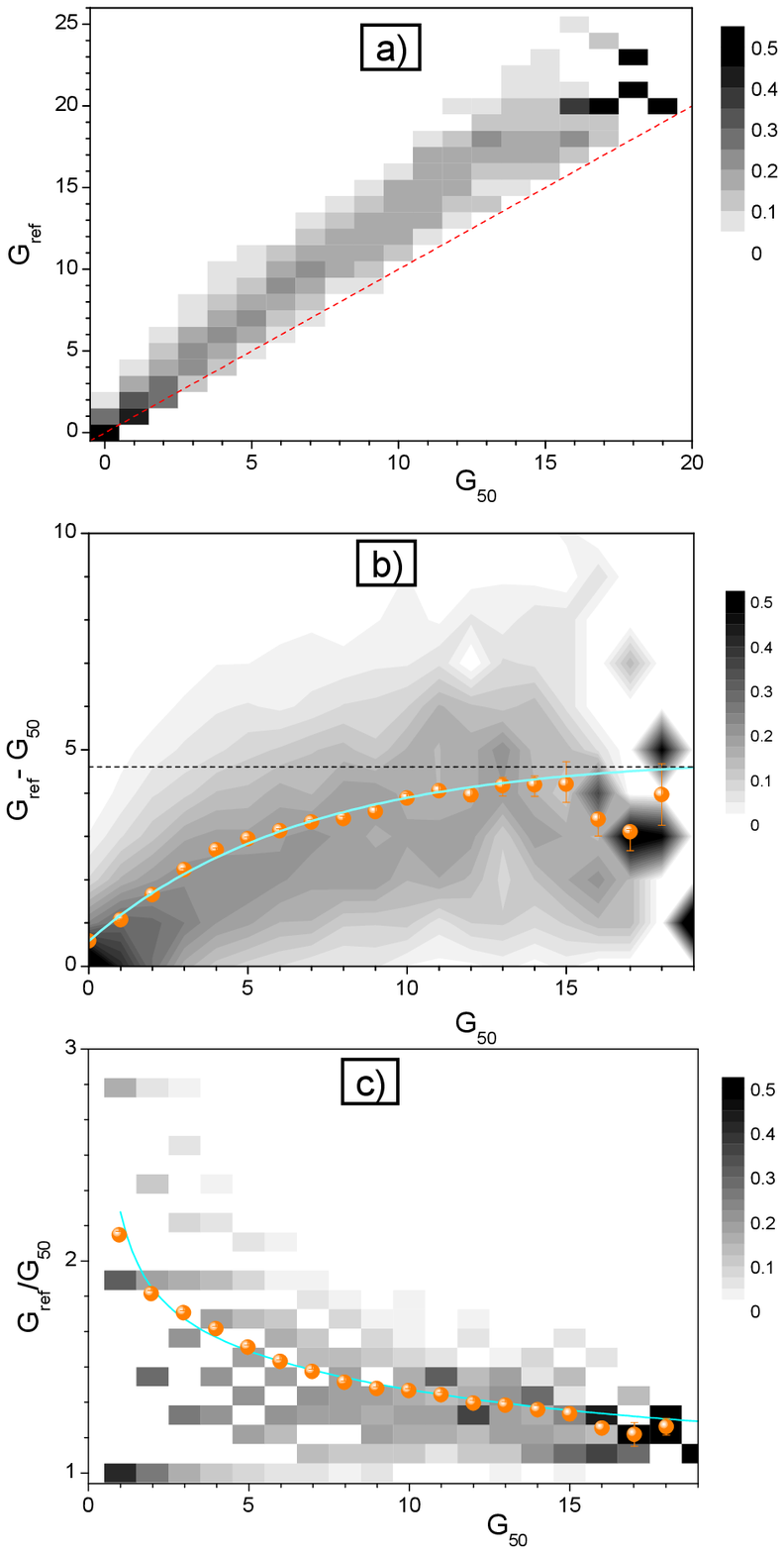}}
\caption{
Relation between daily sunspot-group numbers as counted by a `perfect' observer $G_{\rm ref}$ and by an observer with
 the acuity threshold $A_{\rm th}=50$ msd, $G_{50}$, using data from the reference data set.
Panel a: The PDF of the scatter plot of $G_{\rm ref}$ \textit{vs.} $G_{50}$.
The red dashed line marks the diagonal.
Panel b: The difference ($G_{\rm ref}-G_{50}$) \textit{vs.} $G_{50}$.
The orange dots with error bars depict the mean values (with the asymmetric errors of the mean) of the difference in each
 PDF strip for fixed $G_{50}$.
The cyan line is the best fit Equation (\ref{Eq:exp}), while the horizontal dashed line is the value of $R_\infty$.
Panel c: The correction factor, $c_{50}\equiv G_{\rm ref}/G_{50}$ \textit{vs.} $G_{50}$.
The orange dots with error bars depict the mean values (with the errors of the mean) of the difference in each
 PDF strip for fixed $G_{50}$.
The cyan line corresponds to that in panel b.
The color code for all PDFs is shown on the right.
 }
\label{Fig:rgo50}
\end{figure}

In the subsequent section we study how the fact that the size distribution of sunspot groups changes
 with the level of solar activity affects solar observers of different quality and their mutual
 inter-calibration.

\section{Observer's View}
\label{Sec:obs}

\subsection{``Imperfect'' vs. ``perfect'' observers}
\label{Sec:imp}
We assume (similar to \opencite{lockwood_SP3_2016} and \opencite{usoskin_ADF_16}) that a solar observer is imperfect {and}
 has an observational acuity threshold [$A_{\rm th}$] for the
 (uncorrected for foreshortening) sunspot-group area so that he/she reports all the groups bigger than $A_{\rm th}$
 and miss all the groups smaller than $A_{\rm th}$.
Accordingly, we simulated a set of sunspot-group records produced by pseudo-observers
 synthesized from the reference dataset and characterized only by the acuity threshold [$A_{\rm th}$].

We define the ``correction factor'' $c_A$ as the ratio of the number of sunspot groups $G_{\rm ref}$
 reported by the perfect observer ($A_{\rm th}=0$) to that $G_A$ by an imperfect observer with the
 finite acuity threshold [$A$] ($A_{\rm th}=a$), as a function of $G$:
\begin{equation}
c_A(G_A) \equiv {G_{\rm ref}\over G_A}.
\label{Eq:ca}
\end{equation}
Figure~\ref{Fig:CF} depicts ($c_A$-1) as function of $G_A$ for several values of the threshold $A_{\rm th}$.
One can see that the relative observational errors (the fraction of missed {sunspot groups to the ``true'' number of groups})
 of a ``poor'' observer rapidly decrease with the level of solar activity.
For example, while the observer with $A_{\rm th}=100$ msd would count {roughly} $^2$\hspace{-0.05cm}/\hspace{-0.05cm}$_3$
 less groups at very low activity $G_{100}=1$, the fraction of missed small groups
 is only 20-25\% for the high activity days $G_{100}=15$.
It is obvious that applying a constant correction factor (as used in the $k-$factor method) is inappropriate, {since}
 this would distort the entire series and overcorrect the periods of high solar activity.
Here we introduce the $c-$factor to correct sunspot-group counts by an ``imperfect'' observer to the reference observer.
The concept of the $c-$factor is similar to that of the $k-$factor but depends on the level of solar activity quantified
 as $G_A$.

In Figure~\ref{Fig:rgo50} we depict the relation between the reference and ``poor'' observers for the
 acuity threshold of $A_{\rm th}=50$ msd.
Panel a shows the conversion matrix for the reference observer and the one with $A_{\rm th}$, constructed in the same way
 as by \inlinecite{usoskin_ADF_16}, for all days during the period {1916--1976}.
The matrix represents the statistics of the reported $G-$values for the two observers, normalized to unity in each
 column so that it represents the probability density function (PDF) of the number of groups reported by the reference observer
 $G_{\rm ref}$ for the days when the poor observer reported $G_{50}$ groups.
One can see that the poor observer always counts less groups than the reference one (all shaded areas lie above the diagonal),
 but the shape of the relation is not clear and could seemingly be approximated by a straight line.

To verify our previous assertion, we show in Figure~\ref{Fig:rgo50}b the PDF for the difference between the observers,
 \textit{i.e.} $D_{50}=(G_{\rm ref}-G_{50})$ as a function of $G_{50}$.
One can see that the relation is essentially nonlinear but bends to become flat at high activity.
While the actual spread of the distribution (grey shading) is wide, this feature is clearly visible from the mean
 values (orange dots on the plot) of
 $D_{50}$ that form a smooth curve which tends to reach a saturation at high values of $G_{50}$.
This relation is very smooth and we approximate it with a simple euqation
\begin{equation}
D_A(G_A)=R_{\infty}-(R_\infty-R_0)\, \exp(-\alpha\, G_A),
\label{Eq:exp}
\end{equation}
where $R_0=D_A(0)$ is the mean value of $G_{\rm ref}$ at $G_A=0$, $R_\infty$ is the asymptotic value (shown as
 the horizontal dotted line
 in Fig.~\ref{Fig:rgo50}b), and $\alpha$ defines the exponential rate.
We note that $R_0$ represents the mean number of sunspot groups reported by the reference observer for days
 when the ``poor'' observer reports no sunspot group.
For this particular case, $R_0=0.64_{-0.01}^{+0.03}$ implying that on average the reference observer sees 0.64 sunspot groups
 on days when the ``poor'' observer with $A_{\rm th}=50$ msd reports no groups because of their small size.

Since the value of $R_0$ is fixed for a given observer, Equation (\ref{Eq:exp}) includes two
 free parameters $\alpha$ and $R_\infty$.
We fitted this relation to the mean values of $D_A$ (orange dots in Fig.~\ref{Fig:rgo50}b) using the $\chi^2$ method.
The best-fit parameters for the dependence shown in Fig.~\ref{Fig:rgo50}b are: $\alpha=0.15\pm 0.01$, $R_\infty=4.89\pm 0.22$.
The value of $\chi^2$ for 14 degrees of freedom (DoF) is 15, implying a good fit.

The correction factor $c_A(G_A)$ (see Equation~\ref{Eq:ca}) can be computed from the difference $D_A$ as
\begin{equation}
\label{Eq:cf}
c_A(G_A) \equiv {G_{\rm ref}\over G_A} = {D_A \over G_A}+1,
\label{Eq:CC}
\end{equation}
as shown in Fig.~\ref{Fig:rgo50}C.
The proposed empirical relation (blue curve) describes the dependence of $c_A$ on $G_A$ quite well.
We note that this curve was not fitted again to the data points but simply recalculated from that shown in panel b.
The value of $c_A$ is not defined for $G_A=0$.

\subsection{Empirical Dependence}
\label{Sec:emp}
We have repeated the exercise described in Section~\ref{Sec:imp}, \textit{i.e.} fitting Equation~\ref{Eq:exp}
 to the difference $D_A$, for different values of the observational acuity threshold [$A_{\rm th}$] from 10 to 200 msd.
This range corresponds to actual observers with proper telescopes (\opencite{vaquero09,arlt13,neuhauser15,usoskin_ADF_16}).
Generally, an observer with the visual acuity threshold exceeding 100 msd would be considered as one with very poor quality.

\begin{figure}
\centering \resizebox{\columnwidth}{!}{\includegraphics[bb = 58 326 532 754]{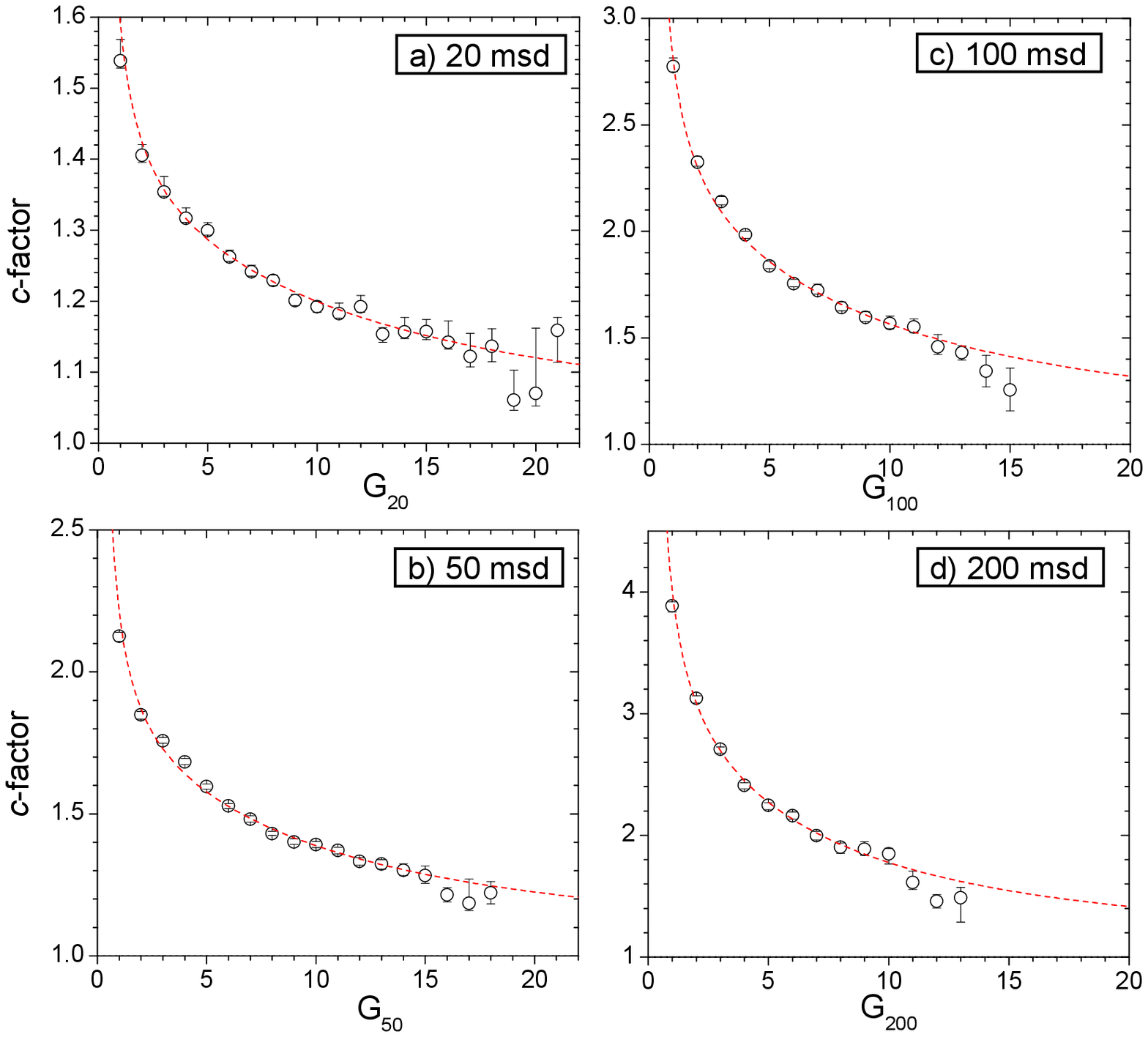}}
\caption{
The correction $c-$factors for different values of the acuity threshold [$A_{\rm th}$] as indicated in the insets.
Only the means with their standard errors are shown (see Fig.~\ref{Fig:rgo50}c) as open circles.
The red dashed curves are the best-fit relations given by Equation~\ref{Eq:cf}.
 }
\label{Fig:fits}
\end{figure}

Some examples of the fits are shown in Figure~\ref{Fig:fits} for several values of $A_{\rm th}$.
Although the fits were actually done for the $D_A(G_A)$ distributions (similar to that shown in Fig.~\ref{Fig:rgo50}b),
 we show here the $c-$factors.
We see that the equation fits the data almost perfectly when the statistic is good but the
 errors bars of the data points increase for higher values of $G_A$ because of a poorer statistic. .
The fit is always good in the entire range of the $A_{\rm th}$ values analyzed here, and the values of $\chi^2$ are
 typically around 1 \textit{per} DoF with a variability between 0.5 and 1.5 \textit{per} DoF, indicating a good agreement.

It is interesting to note that the best-fit curves tend to slightly overestimate the
 $c-$factor for the highest activity periods (see Figure~\ref{Fig:fits}).
This implies that an ``imperfect'' observer appears not as bad as it should be according to its acuity limitation.
However, this tendency is not statistically significant, and we do not consider it in this study.
Since we use the method of minimizing $\chi^2$ to fit the curve, the contribution of these points is small.

\begin{figure}
\centering \resizebox{\columnwidth}{!}{\includegraphics[bb = 63 206 528 794]{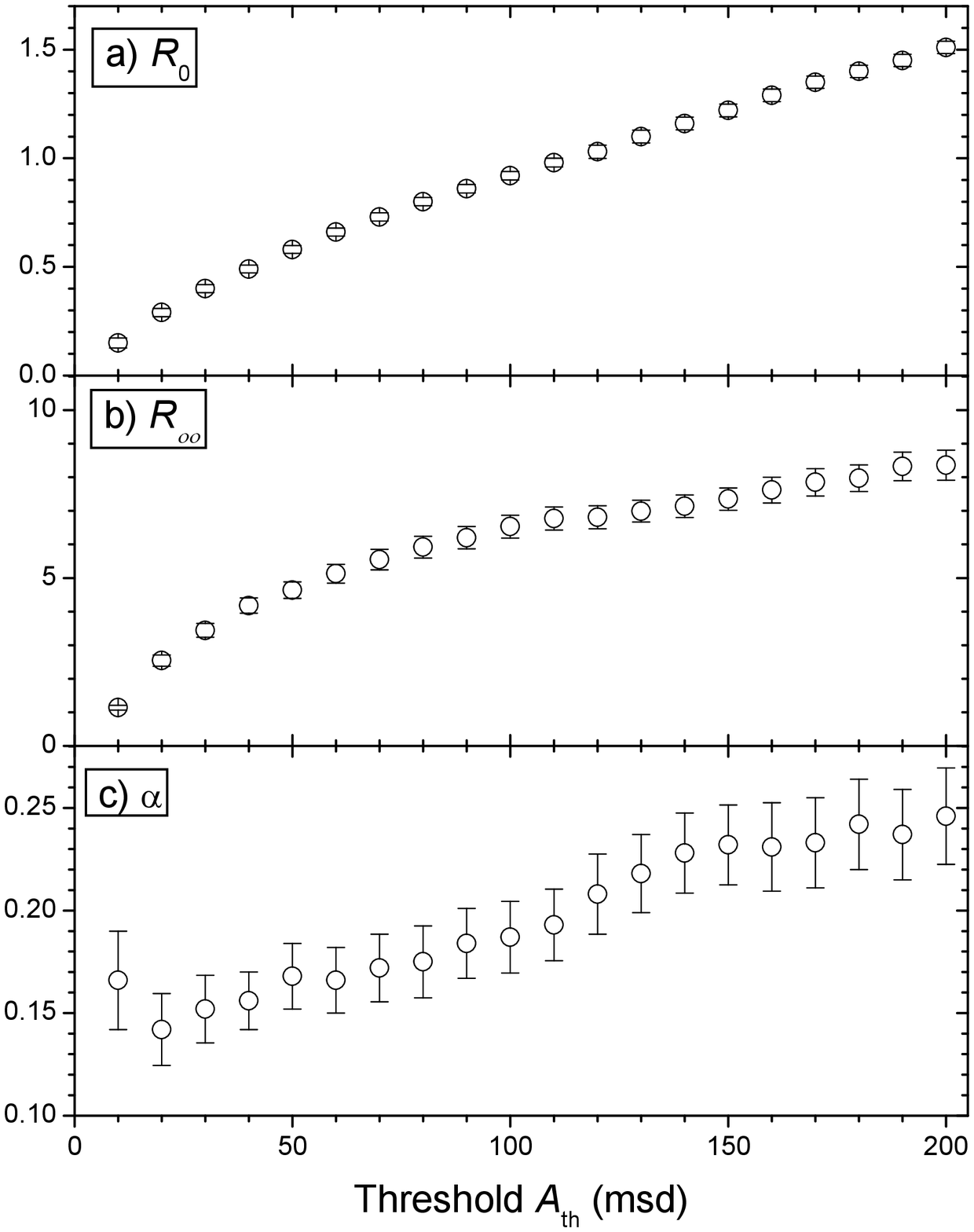}}
\caption{
Dependence of the best-fit parameters of Equation (\ref{Eq:exp}) on the acuity threshold [$A_{\rm th}$], using the
 reference dataset.
The parameters are shown with the 68\% confidence interval ($1\sigma$) uncertainties.
 }
\label{Fig:par}
\end{figure}

The dependence of the parameters of the empirical relation on the value of $A_{\rm th}$ is shown in Figure~\ref{Fig:par}.
One can see a smooth growth of the parameters with increasing acuity threshold.
The parameter $R_0$ raises from its obvious zero value to {nearly} 1.5.
The parameter $R_\infty$ grows from its obvious zero value to
 about eight and tends to saturate there.
The parameter $\alpha$ varies between 0.15 and {0.27}.

The empirical relation studied here (Equations~\ref{Eq:exp}--\ref{Eq:CC}) implies that the fraction of small sunspot groups
 is not constant as a function of solar activity level, but declines exponentially with the activity as is apparent from
 Figure~\ref{Fig:fits}.
Over the solar cycle, the number of sunspot groups increases mostly because of large groups, while the amount of small groups
 remains nearly constant
 in accordance with the analysis presented in Section~\ref{Sec:dist}.

\subsection{A Test: Wolf \textit{vs.} Wolfer}
\label{Sec:WW}

Here we study the relation between the sunspot-group numbers reported by two classical sunspot observers of the
 second half of the 19th century: Rudolf Wolf and Alfred Wolfer, both from Z\"urich Observatory.
They were primary observers for the Wolf (and its successor the International) sunspot number series
 and thus defined the values of the sunspot numbers the late 19th and earlier 20th centuries.
A scaling $k-$ factor of 1.66 is traditionally applied to the sunspot (group) number by Wolf to match it with Wolfer in quality,
 as initially proposed by Wolfer himself.
This factor was used continuously (\opencite{clette14,svalgaard16})
 in the daisy-chain procedure to calibrate the sunspot number series ever since.
Thus, it is important to verify the inter-calibration of the two observers.
We use all the days (4385 in total) when we have observations from both observers for the period 1876--1893.
This analysis is close to that of \inlinecite{usoskin_ADF_16} but is focused on the use of the proposed parameterization
 (Equation (\ref{Eq:exp})).

\begin{figure}
\centering \resizebox{8cm}{!}{\includegraphics[bb = 59 203 335 739]{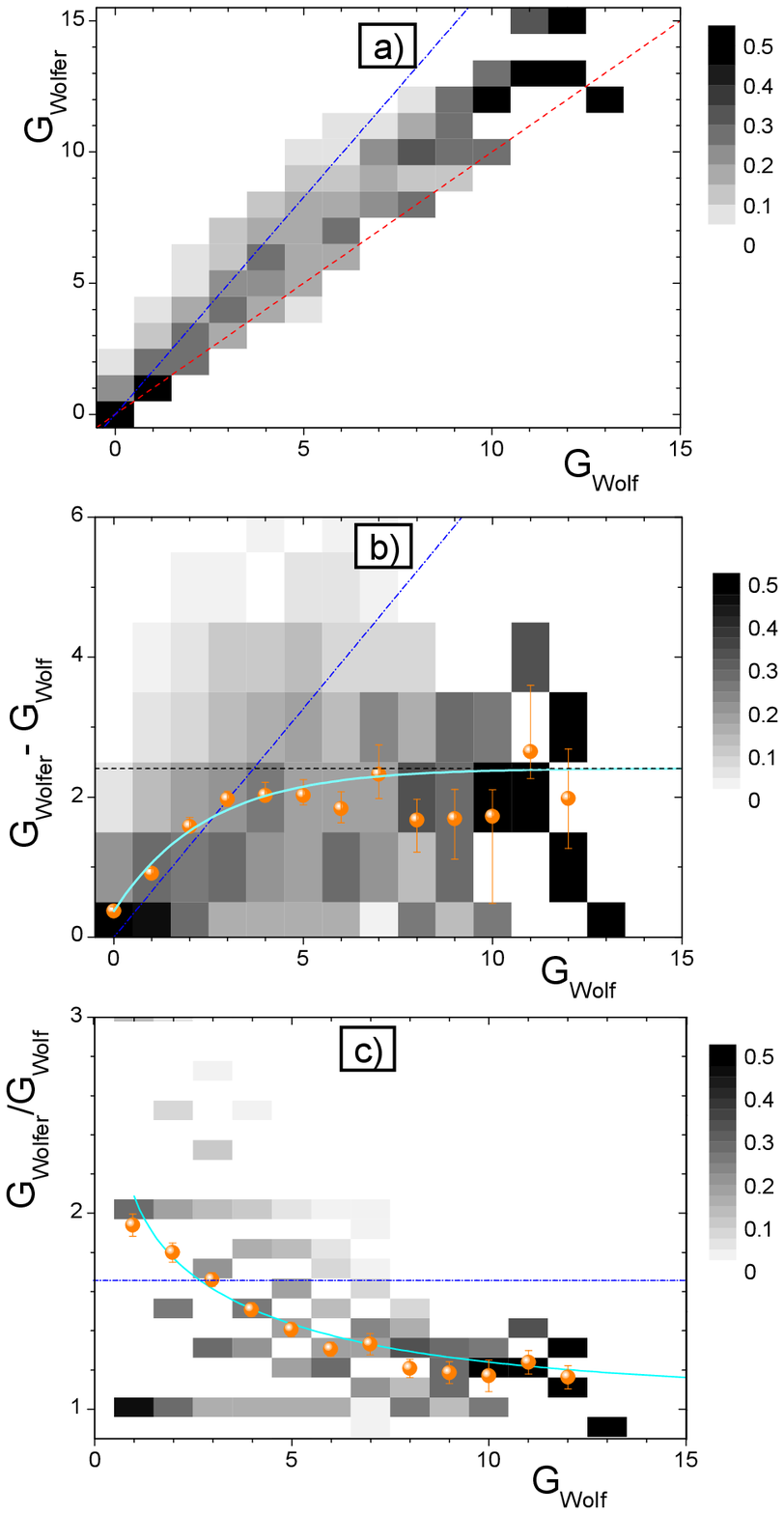}}
\caption{
Similar to Figure~\ref{Fig:rgo50} but for the relation between group numbers as reported by Wolfer, $G_{\rm Wolfer}$, and
 Wolf, $G_{\rm Wolf}$.
The blue dashed lines in all panels denote the constant scaling $k-$fator of 1.66 often used to scale Wolf to Wolfer.
 }
\label{Fig:WW}
\end{figure}

Figure~\ref{Fig:WW}a shows a conversion matrix, constructed in the same way as that in Figure~\ref{Fig:rgo50} but for
 the Wolfer \textit{vs.} Wolf data, so that Wolfer is now the reference observer and Wolf is a ``poor'' one.
One can see that Wolfer was indeed a better observer than Wolf since he reported more groups than Wolf did, for the same days.
This is shown by as the grey area in {Figure~\ref{Fig:WW}a that lies} systematically above the diagonal (red dotted line).
On the other hand, the blue dashed line (the $k-$factor of 1.66) lies systematically above the grey area
 for $G_{\rm Wolf}>4$.

Figure~\ref{Fig:WW}b depicts the difference $G_{\rm Wolfer}-G_{\rm Wolf}$ as function of $G_{\rm Wolf}$, similar to
 Figure~\ref{Fig:rgo50}3.
One can see that the relation between the number of groups reported by the two observers is not linear, but
 again has the same shape of an asymptotic approach to the constant difference.
The value of $R_0$ was found to be $0.401_{-0.13}^{+0.06}$, implying that Wolfer on average reported 0.4 sunspot groups for
 days when Wolf reported none, due to the difference in the instrumentation and eyesight.
Mean values of the distribution (orange balls in the figure) can be fitted well by Equation (\ref{Eq:exp}) with
  $R_\infty=2.43$.
The linear relation ($k-$factor of 1.66, the blue dashed line) does not describe the relation in a reasonable way
 as it systematically overestimates the corrected Wolf's records for days with mid- and high-activity ($G_{\rm Wolf}>4$).

Figure~\ref{Fig:WW}c shows the correction factor ($G_{\rm Wolfer}/G_{\rm Wolf}$) as a function of $G_{\rm Wolf}$.
It is obviously nonlinear: while Wolf underestimated (comparing to Wolfer) the number of groups by a factor of two
 during the low activity periods (one sunspot group \textit{per} day), his under-count of groups was only 20\% for
 the days with a number of groups exceeding eight.
Accordingly, the assumption of the constancy of the correction $k-$factor at the level of 1.66 (the blue dashed line)
 is invalid in this case.

\begin{figure}
\centering \resizebox{\columnwidth}{!}{\includegraphics[bb = 72 525 468 741]{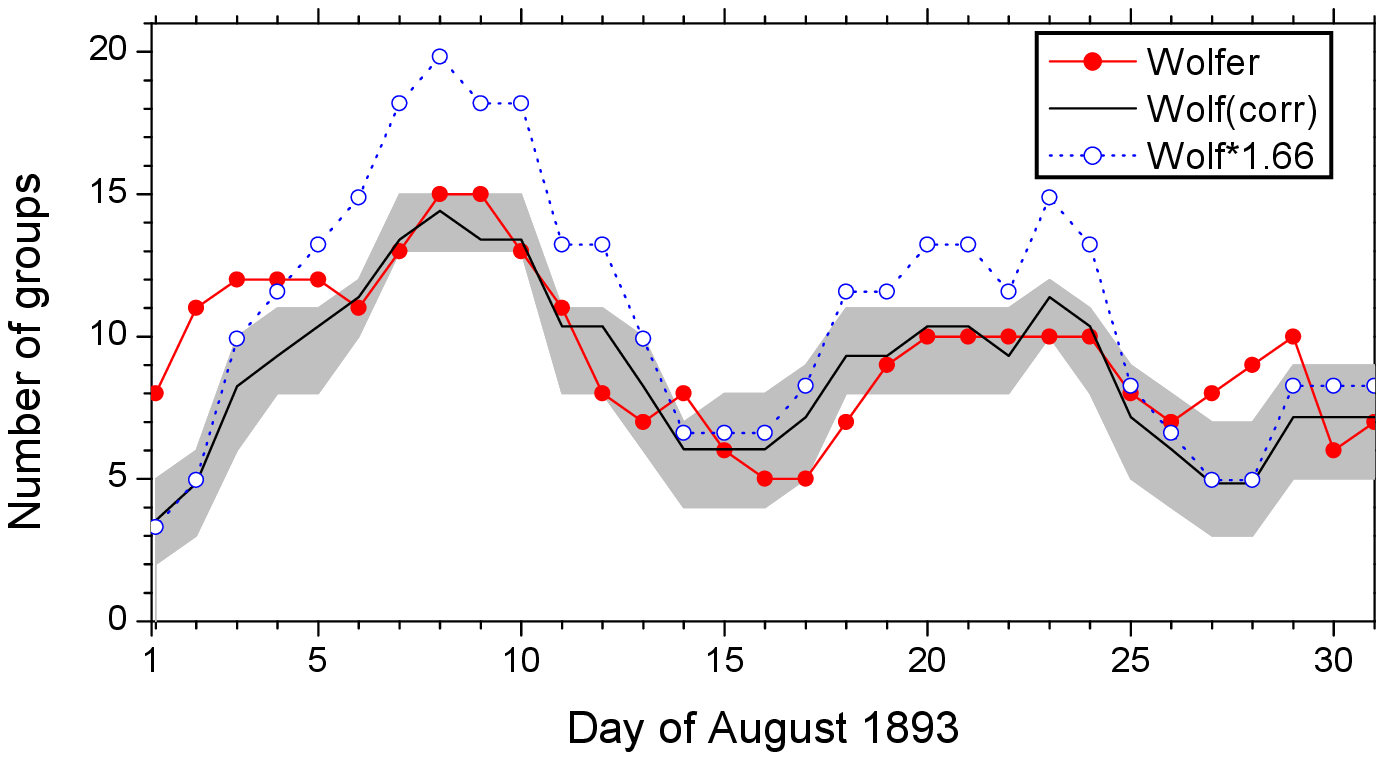}}
\caption{
Daily number of sunspot groups for August 1893.
The red curve depicts the number of groups reported by Wolfer,
the black curve with its 68\% confidence interval denotes the data by Wolf normalized to Wolfer by the method described here,
 and the blue dotted curve is the data by Wolf scaled by the $k-$factor 1.66.
 }
\label{Fig:Aug93}
\end{figure}

A principle assumption behind the $k-$factor methodology (see Section~\ref{Sec:Intro}) is that the relation between the number of sunspots (groups) reported
 by different observers is linear.
However, as shown here for the example of two famous observers, the relation between the number of sunspot groups reported
 by them is essentially nonlinear and, when applying the constant $k-$factor, leads to underestimate of the number of groups
 during low activity but overestimate is during high activity periods.
This is illustrated in Figure~\ref{Fig:Aug93}, where daily group numbers are shown for August 1893 (a month with high activity)
 by both observers: the original group counts by Wolfer, the counts by Wolf corrected using the constant $k-$factor of 1.66,
 and the counts by Wolf corrected using Equation (\ref{Eq:exp}).
We see that for the period from 6th to 26th of August, when the temporal profiles of the number of groups
 by Wolf and Wolfer were close to each other, indicating that they reported the same sunspot activity evolution,
 the correction based on the $k-$factor yields a good agreement
 for the days with moderate activity ($G_{\rm Wolfer}<8$) but systematically overestimates counts
 by 3--4 groups (40--60\%) for the days with high activity ($G_{\rm Wolfer}>10$).
On the other hand, the nonlinear method proposed here reproduces the level for the entire period correctly.

\section{Conclusions}

We have studied the distribution of sunspot group size or area and its dependence on the level of solar activity.
We show that the size distribution of sunspot groups cannot be assumed constant but varies significantly
 with solar activity.
The fraction of small groups, which can be potentially missed by an ``imperfect'' solar observer, is found to be
 not constant but decreasing with the level of solar activity.
An empirical relation (Equation (\ref{Eq:exp})) is proposed which describes the amount of small groups as a function of the
 solar activity level.
It is shown that the number of small groups asymptotically approaches a saturation level so that high solar activity
 is largely defined by {big} groups.

We have studied the effect of the changing sunspot group area on the ability of realistic observers to see
 and report the daily number of sunspot groups.
It is shown that the relation between the numbers of sunspot groups as seen by different
 observers with different observational acuity thresholds (defined by the quality of their instrumentation and
  eye sights) is strongly non-linear and cannot be approximated by a linear scaling, in contrast to how it was traditionally
  done earlier.
We propose to use the observational acuity threshold $A_{\rm th}$ to quantify the quality of each observer, instead of
 the relative $k-$factor used earlier.
The value of $A_{\rm th}$ means that all sunspot groups bigger than $A_{\rm th}$ would be reported
 while all the groups smaller than $A_{\rm th}$ missed by the observer.
We have introduced the non-linear $c-$factor, based on the observer's acuity threshold $A_{\rm th}$, which can be used
  to correct each observer to the reference conditions.

The method has been applied to a pair of principal solar observers of the 19th century, Rudolf Wolf and Alfred Wolfer
 of Z\"urich.
We have shown that the earlier used linear method to correct Wolf data to the conditions of Wolfer, using the constant $k-$factor
 of 1.66, tends to overestimate of solar activity around solar maxima.

This {result} presents a new tool to recalibrate different solar observers to the reference conditions.
A full recalibration based on the new method will be a subject of a forthcoming work.

\section*{Acknowledgements}
IGU and GAK acknowledge support by the Academy of Finland to the ReSoLVE Center of Excellence (project N$^0$. 272157).
T.C. acknowledges the postgraduate fellowship of the International Max Planck Research School on Physical Processes in the Solar System and Beyond.

\section*{Disclosure of Potential Conflicts of Interest}
The authors declare that they have no conflicts of interest.


\end{article}
\end{document}